\documentclass[]{aastex631}

\newcommand{\sinc}{\mathrm{sinc}}
\shorttitle{{ A Coronagraph with a Sub-$\lambda/D$ IWA and a Moderate Spectral Bandwidth}}
\shortauthors{Itoh \& Matsuo.}
\usepackage{comment}

\graphicspath{{./}{figures/}}

\begin{document}
\title{{ A Coronagraph with a Sub-$\lambda/D$ IWA and a Moderate Spectral Bandwidth}}

\author[0000-0003-2690-7092]{Satoshi Itoh}
\author[0000-0001-7694-5885]{Taro Matsuo}
\affiliation{ Department of Particle and Astrophysics,\\ Graduate School of Science, Nagoya University,\\ Furocho, Chikusa-ku, Nagoya, Aichi, 466-8602, Japan}



\begin{abstract}
{ Future high-contrast imaging spectroscopy with a large segmented telescope will be able to detect atmospheric molecules of Earth-like planets around G- or K-type main-sequence stars.
Increasing the number of target planets will require a coronagraph with a small inner working angle (IWA), and wide spectral bandwidth is required if we enhance a variety of detectable atmospheric molecules.  
 To satisfy these requirements, in this paper, we present a coronagraphic system that provides an IWA less than 1$\lambda_0 / D$ over a moderate wavelength band, where  $\lambda_0$ is the design-center wavelength  and $D$ denotes the full-width of the rectangular aperture included in the telescope aperture.
A performance simulation shows that the proposed system approximately achieves a contrast below $10^{-10}$ at 1$\lambda_0 / D$ over the wavelengths of 650--750nm.
In addition, this system has a core throughput $\geq$ 10\% at input separation angles of $\sim$ 0.7--1.4$\lambda_0 / D$; { to reduce telescope time, we need prior information on the target's orbit by other observational methods to a precision higher than the width of the field of view.}
For some types of aberration including tilt aberration, the proposed system has a sensitivity less than ever-proposed coronagraphs that have IWAs of approximately $1\lambda_0/D$.
In future observations of Earth-like planets, the proposed coronagraphic system may serve as a supplementary coronagraphic system dedicated to achieving an extremely small IWA.}
\end{abstract}



\section{Introduction} \label{sec:intro}
The spectral characterization of the atmosphere of an Earth-like planet enables us to investigate its habitability and the potential  existence of  biosignatures on the planetary surface   \citep[e.g.,][]{DesMarais+2002,Seager+2016,Kaltenegger+2017,Fujii+2018}.
A broad spectral bandwidth is critical for detecting various atmospheric molecules.
Although the bandwidth can be widened relatively easily for transit spectroscopy \citep[e.g.,][]{Tsiaras+2019,Benneke+2019}, it is relatively challenging for high-contrast imaging spectroscopy, which is necessary for observations of Earth-like planets around G- or K-type main-sequence stars; currently available wavefront compensation and coronagraph masks limit the spectral bandwidth of a coronagraph system to 10\%--20\% \citep{Trauger+2012,Cady+2017}.

Future large telescopes including ground-based extremely large telescopes (ELTs) and the large UV optical infrared space telescope concept will be built with segmented mirrors. This is because a large primary mirror is required to reduce the impact of zodiacal light on the signal-to-noise ratio and to observe more distant planets \citep[e.g.,][]{Kasting+2009}.
We thus need to achieve extremely high-contrast for a complicated pupil with segment gaps and occultation by the secondary-mirror-support structure. \citep[e.g.,][]{Guyon+2010,Guyon+2014,Mawet+2011a,Pueyo+2013}.
The use of an apodized pupil Lyot coronagraph with a binary mask \citep{N'Diaye+2016} and a vector vortex coronagraph with deformable mirrors and pupil apodization \citep[e.g.,][]{2018JATIS...4a5004R} are promising approaches for on- and off-axis segmented telescopes, respectively.

Nevertheless, the inner working angles (IWAs) of these methods limit the number of potentially habitable planets observable with possible future space telescopes \citep[][]{Stark+2019}.
The coronagraphs for the LUVOIR-A and -B mission concepts can achieve  IWAs of about 3.7 and 2.5 times the diffraction limit, respectively; { these IWAs are adopted mainly based on the trade-off studies between the  the stellar-light leakage due to the realistic aberration budgets of the optical system and IWAs.} 
Consequently, the yield of potentially habitable planets can be increased by achieving high-contrast imaging at a small angular separation from the central stars (less than 2 times the diffraction limit) with a moderate spectral bandwidth. For example, when we observe a planet separated by 1AU from its host star at a wavelength of 700nm with a 6-m diameter telescope, an IWA of 2.5 times the diffraction limit enables the observation of a planet located up to about the 16.6-pc distance. 
In the same condition, for the IWA of 0.8 times the diffraction limit, the distance at which we can observe the planetary system extends up to 52.0pc in principle.

A phase-shift mask on the focal plane and an optimized pupil apodization produce a very small IWA for an arbitrary pupil \citep{Guyon+2010,Guyon+2014}; the phase-shift mask transmits the light outside of a specific radius (typically half the radius of the Airy disk) and gives a $\pi$-radian phase shift to the light inside the radius so that the stellar Airy disk interferes with itself destructively. 
However, because the radius of the phase-changing points must be proportional to the wavelength, the spectral bandwidth of this approach is severely limited.  
Furthermore, because it has a second-order response (second-order null) to low-order aberrations, this type of coronagraph system reacts sensitively to the finite stellar angular diameter and telescope pointing jitter \citep{Belikov+2018}.

\citet{Itoh+2020} have presented a method for achieving a diffraction-limited IWA and the fourth-order response (fourth-order null) to low-order aberrations simultaneously for a non-apodized, segmented pupil; we refer to this method hereafter as the ``IM-2020 system.'' 
It uses a one-dimensional focal plane mask with amplitude modulation and $\pi$-radian phase shift, as well as an optimized Lyot stop; hereafter, we refer to this focal-plane mask as the ``IM-2020 mask.''
We can manufacture this type of focal plane mask that takes values from -1 to 1 achromatically by putting a half-wave plate with spatially variant fast axes between two linear polarizers orthogonal to each other. 
{ This configuration brings simultaneous modulation of amplitudes and $\pi$-radian phases.
} 
Although a single set consisting of an IM-2020 mask and a Lyot stop has a second-order response to low-order aberrations, we can achieve a fourth-order null in a diagonal direction on the focal plane by successively placing two sets orthogonal to each other. 
However, the IM-2020 system allows only a narrow-band usage (0.3\% bandwidth for a contrast of $10^{-10}$). Focusing on the one-dimensional nature of the IM-2020 mask, \citet{Matsuo+2021} upgraded the IM-2020 system to provide a wide-band coronagraph system (spectroscopic fourth-order coronagraphy). 
In spectroscopic fourth-order coronagraphy, we need two diffraction gratings---before and after an optimized focal-plane mask---to disperse the point spread functions (PSFs) perpendicular to the direction of the mask modulation and obtain a white-light pupil from the dispersed PSFs.
The optimized focal-plane mask has an almost one-dimensional pattern, but its scale varies along the direction orthogonal to the mask modulation. 

In this study, we upgrade the IM-2020 system, focusing on the property of the coronagraph that generates ``flat'' leak amplitudes at the Lyot stop because of the increase in the spectral bandwidth.
Here, the ``flat'' amplitude is a complex amplitude that is constant from the perspective of both absolute value and phase.
For light with a wavelength of $\lambda$, the amplitude of the pupil-plane leak from a single set consisting of a mask and a Lyot stop is approximately $(\lambda-\lambda_{0})/\lambda$, where $\lambda_0$ denotes the design-center wavelength. Importantly, $n$ successive sets of the mask and Lyot stop reduce the leak amplitude to $\lbrace(\lambda-\lambda_{0})/\lambda\rbrace^{n}$.
This occurs because a constant leak at the Lyot stop serves as an on-axis point source for the next set of the mask and Lyot stop.
However, a single set of an IM-2020 mask and a Lyot stop unfortunately has approximately 50\% of the off-axis peak throughput, which critically reduces the throughput of its successive sets. 
To overcome this problem, we defined new focal plane masks by adding some periodic modulations to the IM-2020 mask. 
 
In Section 2, we show how to enlarge the effective bandwidth of a coronagraph system with a small IWA by introducing new focal-plane masks.
In Section 3, we simulate the performance metrics of the resulting coronagraphic system.
We summarize the findings of this study in Section 4.

\section{Theory} \label{sec:theory}
\subsection{Problem}
The IM-2020 mask produces a spatially constant complex amplitude at  the Lyot stop as a leak due to wavelength deviation \citep{Matsuo+2021}. 
This leak is proportional to the amplitude of the on-axis source, and thus it can be removed by inserting an IM-2020 system following the leak-producing system.
However, a single set of the IM-2020 mask and Lyot stop unfortunately has an off-axis peak throughput of approximately 50\%, which critically reduces the throughput of its successive sets.
This low off-axis transmittance comes from the following functional form of the IM-2020 mask:\footnote{This coronagraph mask has no wavelength dependence in its physical structure, but the mask function $M_{\zeta}(x_{\lambda})$ depends on the wavelength because of the normalization of $x_{\lambda}$.}
 \begin{equation}
     M_{\zeta}(x_{\lambda})=C_{\zeta}\left(1-2\zeta^{-1}\sinc\left( \frac{2 x_{\lambda} \lambda}{\lambda_0}\right)\right),
 \end{equation}
where $\sinc(x)=\frac{\sin(\pi x )}{\pi x}$; $x_{\lambda}$ is the focal-plane angular coordinate normalized by $\lambda/D$; $D$ is the  pupil diameter; $\lambda$ and $\lambda_0$ respectively denote the observation and design-center wavelengths. 
The factor $\zeta$ works to compensate for the existence of pupil gaps, defined as the ratio of the unobscured region to the entire pupil along one dimension. 
The constant $C_{\zeta}$ adjusts the maximum absolute values of $ M_{\zeta}(x_{\lambda})$ so that it does not exceed unity.
Hence, $C_{\zeta}$ must take a value less than or equal to 0.697, which is the maximum value obtained when $\zeta = 1$.
Consequently, the values of $C_{\zeta}$ reduce the off-axis peak throughput to less than about 50\%. 

\subsection{Solution}
Fortunately, the following focal-plane  mask function relaxes the limit for the off-axis peak throughput of the IM-2020 system: 
\begin{equation}
    M'_{\zeta}(x_{\lambda})=C_{\zeta}'\left(1-2\zeta^{-1}\sinc\left(\frac{4 x_{\lambda} \lambda}{\lambda_0}\right)\right)\cos\left( \frac{\pi x_{\lambda} \lambda}{\lambda_0}\right).
    \label{e1}
\end{equation}
Note that the constant $C_{\zeta}'$ can take values up to 0.900.
The focal plane mask  $M'_{\zeta}(x_{\lambda})$ works like $M_{\zeta}(x_{\lambda})$ for point sources that have separation angles of $N\lambda_{0}/D$ ($N$: an integer)\footnote{The set of functions $\lbrace\mathrm{rect}\left[\alpha \right]e^{2\pi i N\alpha}|N\mathrm{\ is\ an\ integer}.\rbrace$ is a set of orthonormal functions defined on the rectangular pupil $P(\alpha)=\mathrm{rect}\left[\alpha \right]$. 
Since the Fourier transform is a unitary transform, the set of Fourier conjugates $\lbrace\mathrm{sinc}\left(x-N \right)|N\mathrm{\ is\ an\ integer}.\rbrace$, which are diffracted amplitudes of point sources that have separation angles $N\lambda_{0}/D$, is also a set of orthonormal functions. Thus, a linear combination of these functions exists and can be used to express any amplitude spread function diffracted from the rectangular pupil.} although the cosine factor in this mask sharpens these amplitude spread functions (ASFs) to half the widths of the original ones (see Appendix A). The periodic modulation pattern does not produce any additional impact on the pointing-jitter insensitivity and off-axis peak throughput. 
However, the cosine factor that sharpens the ASF degrades the total-energy throughput.

To pursue a high off-axis total-energy throughput, we here optimize the mask modulation pattern; we change the wave shape while keeping its period\footnote{More precisely, we change the functional form to keep the position of the zero point of the function unchanged.}.
The modified mask function is as follows:\footnote{$\mathrm{sgn}(x)= -1\ (x<0),\ 0\ (x=0),\ \mathrm{and} \ 1\ (0<x).$}
\begin{equation}
    M''_{\zeta}(G,H;x_{\lambda})=\sqrt{1-\left[1-\left \lbrace C_{\zeta}'\left(1-2\zeta^{-1}\sinc\left(\frac{4x_{\lambda} \lambda}{\lambda_0}\right)\right)\left|\cos\left( \frac{\pi x_{\lambda} \lambda}{\lambda_0}\right)\right|^H\right \rbrace^2\right]^G}\left(1-2\mathrm{rect}\left[\frac{x_{\lambda} \lambda}{x_0\lambda_0}\right]\right)\mathrm{sgn}\left(\cos\left( \frac{\pi x_{\lambda} \lambda}{\lambda_0}\right)\right),
    \label{e2}
\end{equation}
where $G$ and $H$ are parameters that determine the extent of the modification ($G, H > 0$), and $x_0$ is a mathematical constant that satisfies the equation of $M(2x_0)=0$.
When $G=H=1$, $M''_{\zeta}(G,H;x_{\lambda})$ matches $M'_{\zeta}(x_{\lambda})$. 
Large $G$ and small $H$ produce a rectangular-wave-like off-axis periodic pattern, which results in enhanced total-energy throughput.
When we tune the parameters $G$ and $H$, we need to consider at least  a trade-off relation between  the throughput and amount of the stellar leak.

\section{Performance Simulation} \label{sec:sim}
Here, we simulate the throughput and amount of leak for the proposed  system using the mask function $M''_{\zeta}(G,H;x_{\lambda})$.
{ This simulation utilizes numerical Fourier transform based on the descried Fourier transform (DFT) method; in Appendix B, we show the detail of the numerical calculation and explain its accuracy.} 
\subsection{Setup} \label{subsec:setup}
\begin{figure}[h]
    \centering
    \includegraphics[width=0.8\linewidth]{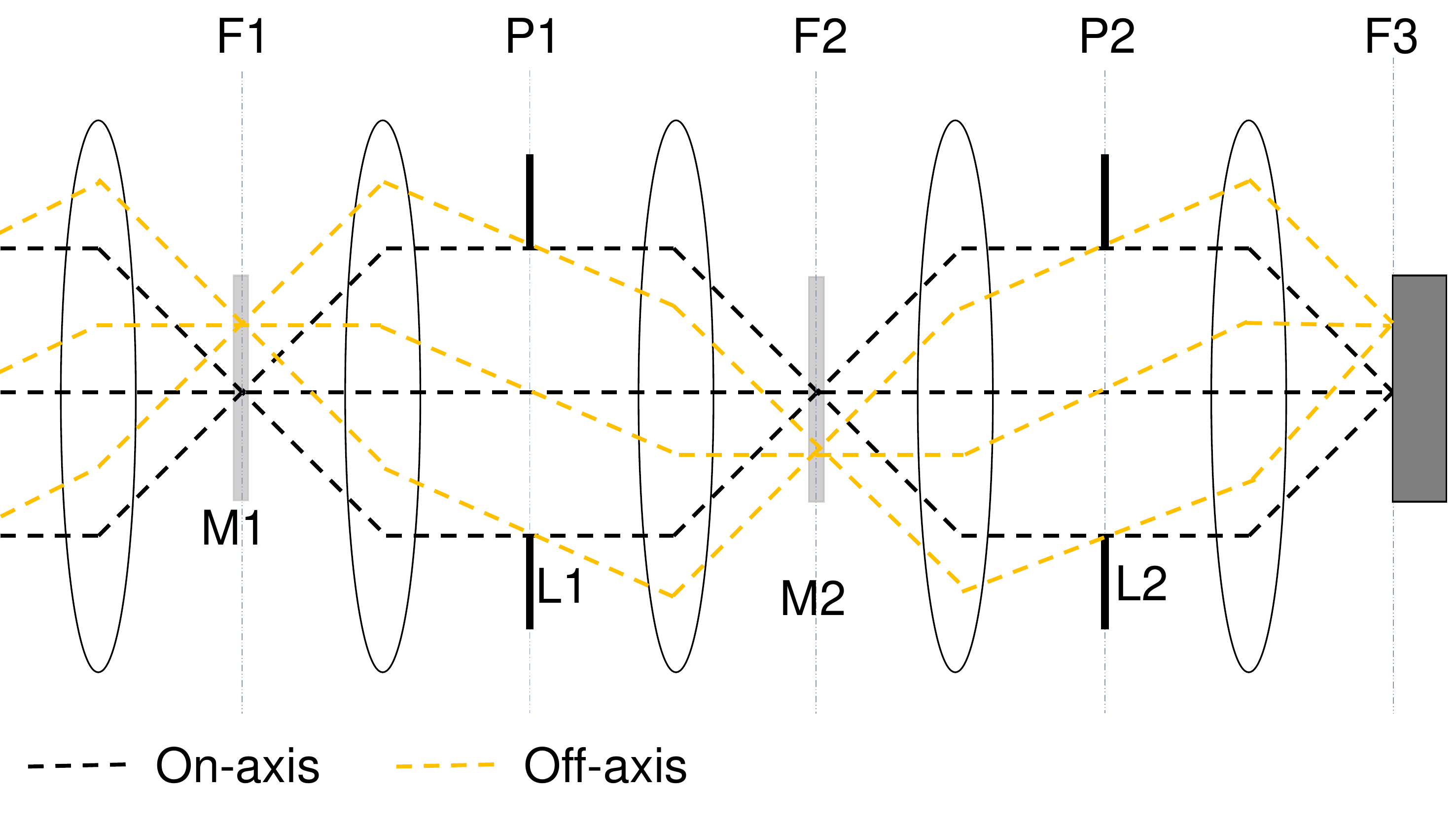}
    \caption{The optical layout for the performance simulation. The black and orange lines represent the light paths of  on- and off-axis point sources, respectively. The abbreviations shown here have the following definitions: the first focal plane (F1), the first focal-plane mask (M1), the first pupil plane (P1), the first Lyot stop (L1), the second focal plane (F2), the second focal-plane mask (M2), the second pupil plane (P2), the second Lyot stop (L2), and the third (final) focal plane (F3). }
    \label{fig:layout}
\end{figure}

\begin{figure}[h]
    \centering
    \includegraphics[width=1.0\linewidth]{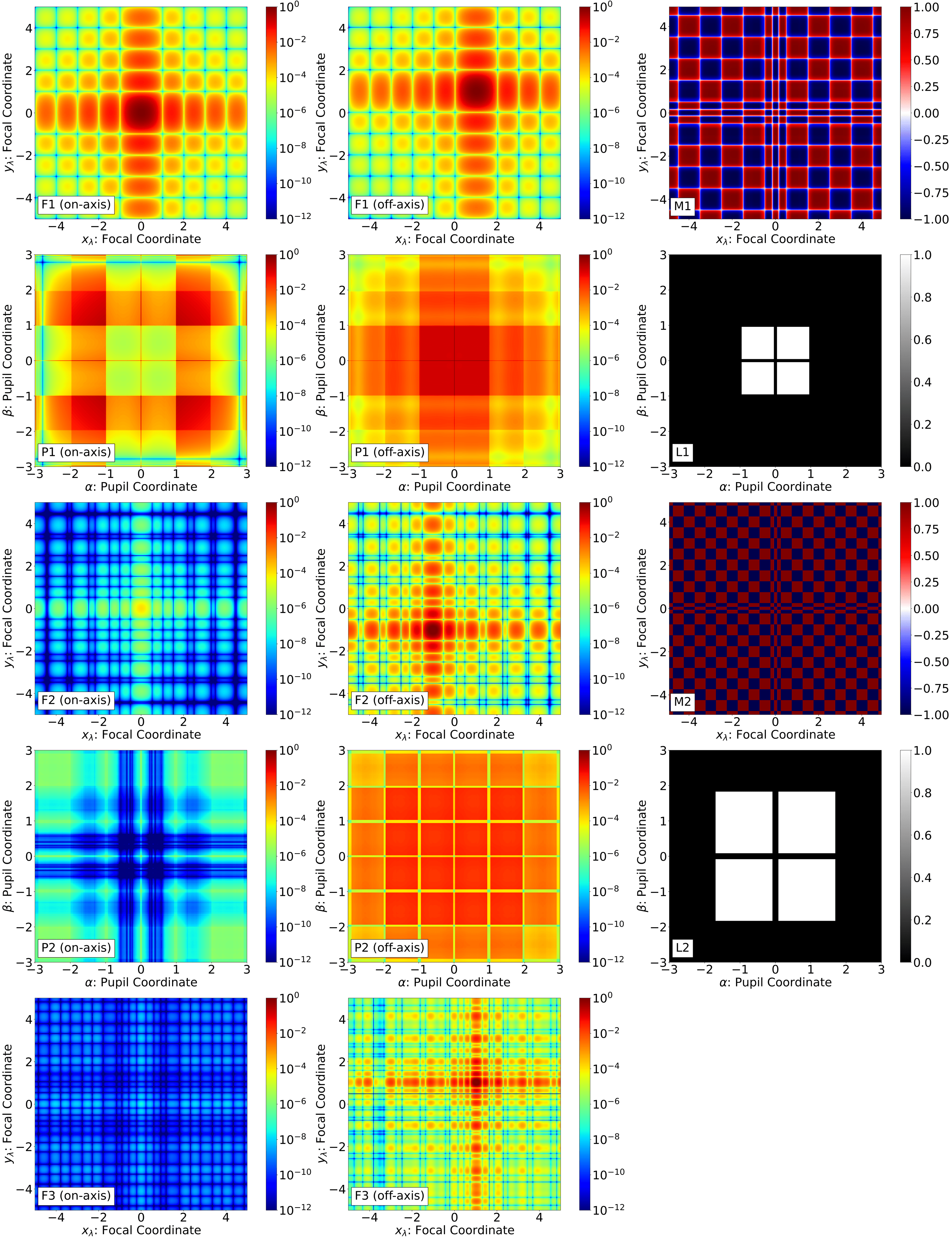}
    \caption{Left column: the normalized intensities of light observed on the focal and pupil planes coming from a point source on the optical axis. Center column: the normalized intensities of light from a point source at the separation angles of $1.0\lambda_0/D$ from the optical axis. Right column: the mask and Lyot-stop functions used in the performance simulation. The abbreviations shown here are the same as in Figure \ref{fig:layout}. The peak intensity at F1 is used to normalize all the intensities.}
    \label{fig:2d}
\end{figure}

\begin{figure}[h]
    \centering
    \includegraphics[width=1.0\linewidth]{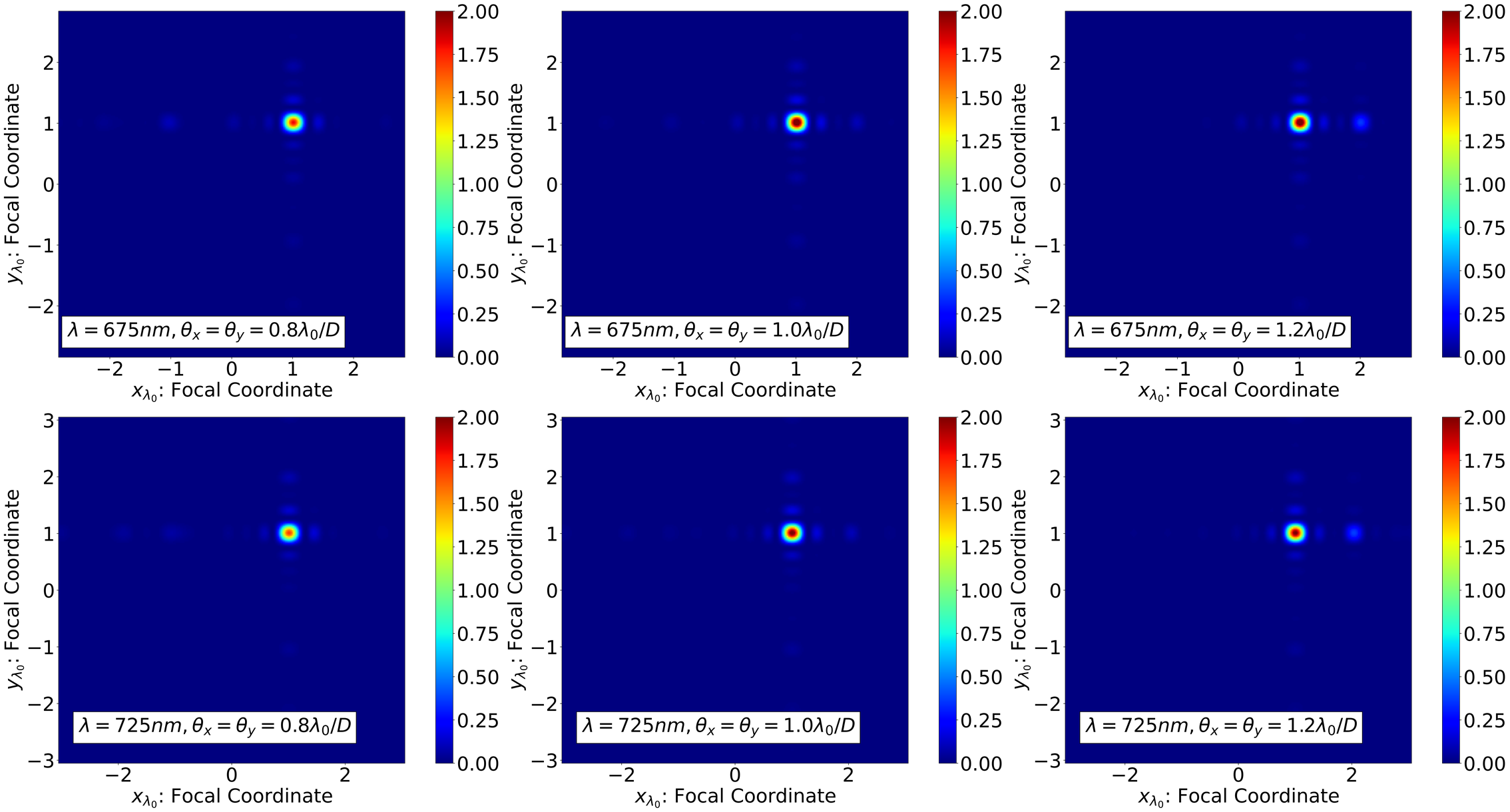}
    \caption{The final output PSFs of input monochromatic point sources at separation angles of $0.8\lambda_0/D$ (left), $1.0\lambda_0/D$ (center) and $1.2\lambda_0/D$ (right) for two observation wavelengths 675nm (top) and 725nm (bottom). Here, the $x$ and $y$ coordinates are normalized by $\lambda_0/D$, regardless of the observation wavelength. The peak locations of the PSFs shown here are approximately independent of both the angular separation and the observation wavelength. }
    \label{fig:inout}
\end{figure}

This simulation uses two successive sets of mask and Lyot stop (Figure \ref{fig:layout}).
We looked for the optimized  parameters via the trial and error method  rather than by using a theoretical algorithm. The design-center wavelengths $\lambda_0$ of these masks are all 712.5nm. The first and second mask functions each have a two-dimensional modulation pattern and are prepared as follows:
\begin{equation}
    M^{\mathrm{2D}}_{k}(x_{\lambda},y_{\lambda})=M''_{\zeta_{k}}\left(G_k,H_k;kx_{\lambda}\right)M''_{\zeta_{k}}\left(G_k,H_k;ky_{\lambda}\right)
\end{equation}
where the integer $k$ takes the value $1$ or $2$, representing the first and second masks, respectively; $\zeta_{1}=1.000$, $\zeta_{2} = 0.956$, $G_1=4$, $H_1=1/2$, $G_2=103$ and $H_2=2/103$. 
We employ the following  first ($k=1$) and second ($k=2$) Lyot stops\footnote{The pupil coordinates $\alpha$ and $\beta$ are normalized by the pupil diameter $D$; the aperture functions of the Lyot stops do not depend on wavelength $\lambda$.}:
\begin{equation}
    L^{\mathrm{2D}}_{k}(\alpha,\beta)= \left(\mathrm{rect}\left[\frac{\alpha}{W_{xk}}\right]-\mathrm{rect}\left[\frac{\alpha}{w_{xk}}\right]\right)\left(\mathrm{rect}\left[\frac{\beta}{W_{yk}}\right]-\mathrm{rect}\left[\frac{\beta}{w_{yk}}\right]\right),
\end{equation}
where $W_{x1}=W_{y1}=1.912$, $w_{x1}=w_{y1}=0.088$, $W_{x2}=3.388$, $W_{y2}=3.657$ and $w_{x2}=w_{y2}=0.168$.  
Figure \ref{fig:2d} shows on- and off-axis, monochromatic light intensities ($\lambda=$700nm)  on the focal and pupil planes together with the mask and Lyot-stop functions used in the performance simulation. Figure \ref{fig:inout} demonstrates the output PSFs of input point sources at three different separation angles from the host star and at two different wavelengths.

We use the ``core throughput'' and ``averaged leak'' as performance metrics. To define these quantities, we use the normalized light power in the region $S=\left\lbrace(x_{\lambda_0},y_{\lambda_0})|0.5\leq x_{\lambda_0} \leq 1.5\ \mathrm{and}\ 0.5\leq y_{\lambda_0} \leq 1.5\right\rbrace$ on the third (final) focal plane; here, the light powers are normalized by the light powers in the PSF main lobes on the first focal plane. 
The averaged leak  and core throughput  are the normalized light power in $S$ for the on-axis source and for the source at the separation angle $1.0\lambda_0/D$, respectively.

\subsection{Results\label{subsec:result}} 
\begin{figure}[h]
    \centering
    \includegraphics[width=1.0\linewidth]{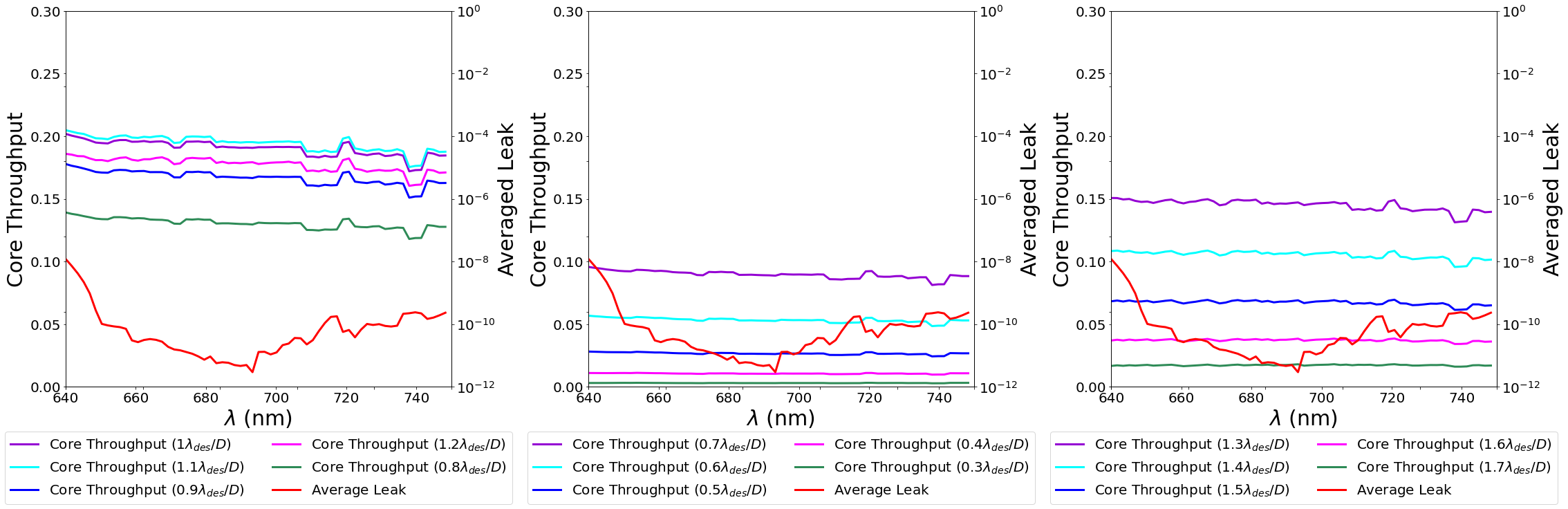}
    \caption{Simulation of the core throughput and averaged leak (defined in Section \ref{subsec:setup}) for different wavelengths. The red curves represent the averaged leak (the right vertical axes). The different-colored curves in each panel represent the core throughput for different separation angles of off-axis sources from a host star (the left vertical axes): $0.8$--$1.2\lambda_0/D$ (left), $0.3$--$0.7\lambda_0/D$ (center) and $1.3$--$1.7\lambda_0/D$ (right). The main reason for the decline in the core throughput for separation angles greater than $1.5\lambda_0/D$ is because the main peak of the PSF is no longer located at 1.0$\lambda_0/D$ but 2.0$\lambda_0/D$. }
    \label{fig:ctal}
\end{figure}

\begin{figure}[h]
    \centering
    \includegraphics[width=0.4\linewidth]{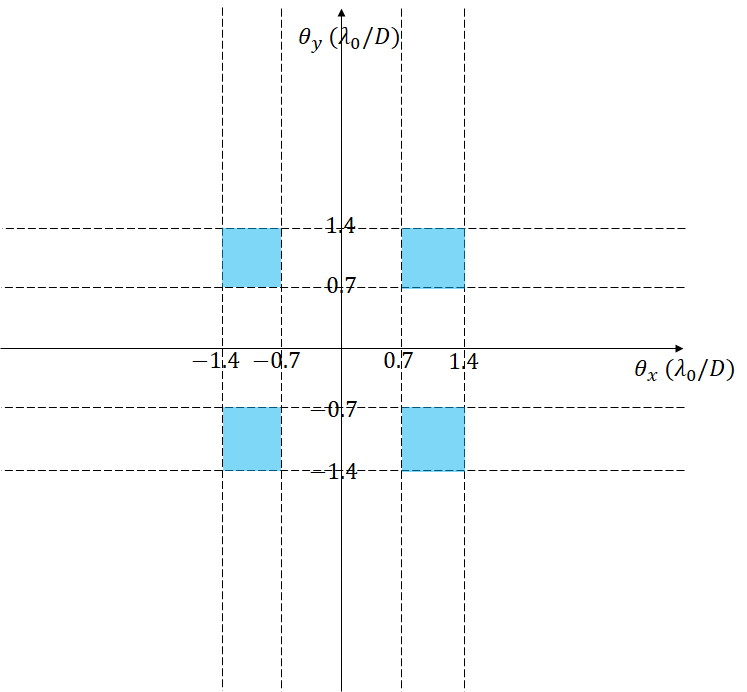}
    \caption{{ Schematic for the region where we can achieve a core throughput higher than 10\% and a contrast of $10^{-10}$ (blue filled areas).  This schematic reflects the results of the simulation shown in Figure \ref{fig:ctal}. Each region has a width of about 0.7$\lambda_0/D$. To save the telescope time, we need to know the target's orbit by other observational methods to a precision higher than 0.7$\lambda_0/D$ before observation.} }
    \label{fig:region}
\end{figure}

Figure \ref{fig:ctal} shows simulations of the core throughput for various off-axes angles and of the averaged leak at 1$\lambda_0/D$. 
The upgraded coronagraph system reduces the averaged leak below approximately $10^{-10}$ over the spectral band of 650--750nm. 
 Further, we find that the core throughput exceeds approximately 10\% for separation angles 0.7--1.4$\lambda_0/D$ over a spectral bandwidth wider than 650--750nm. 
 {
 Figure \ref{fig:region} shows a conceptual diagram for the region where we can achieve a core throughput higher than 10\% and contrast of $10^{-10}$. 
 What is favorable for us to reduce telescope time is to have information on the target's orbit by other observational methods to a precision higher than the width of the field of view 0.7$\lambda_0/D$ before the observation.}
 
\begin{figure}[h]
    \centering
    \includegraphics[width=1.0\linewidth]{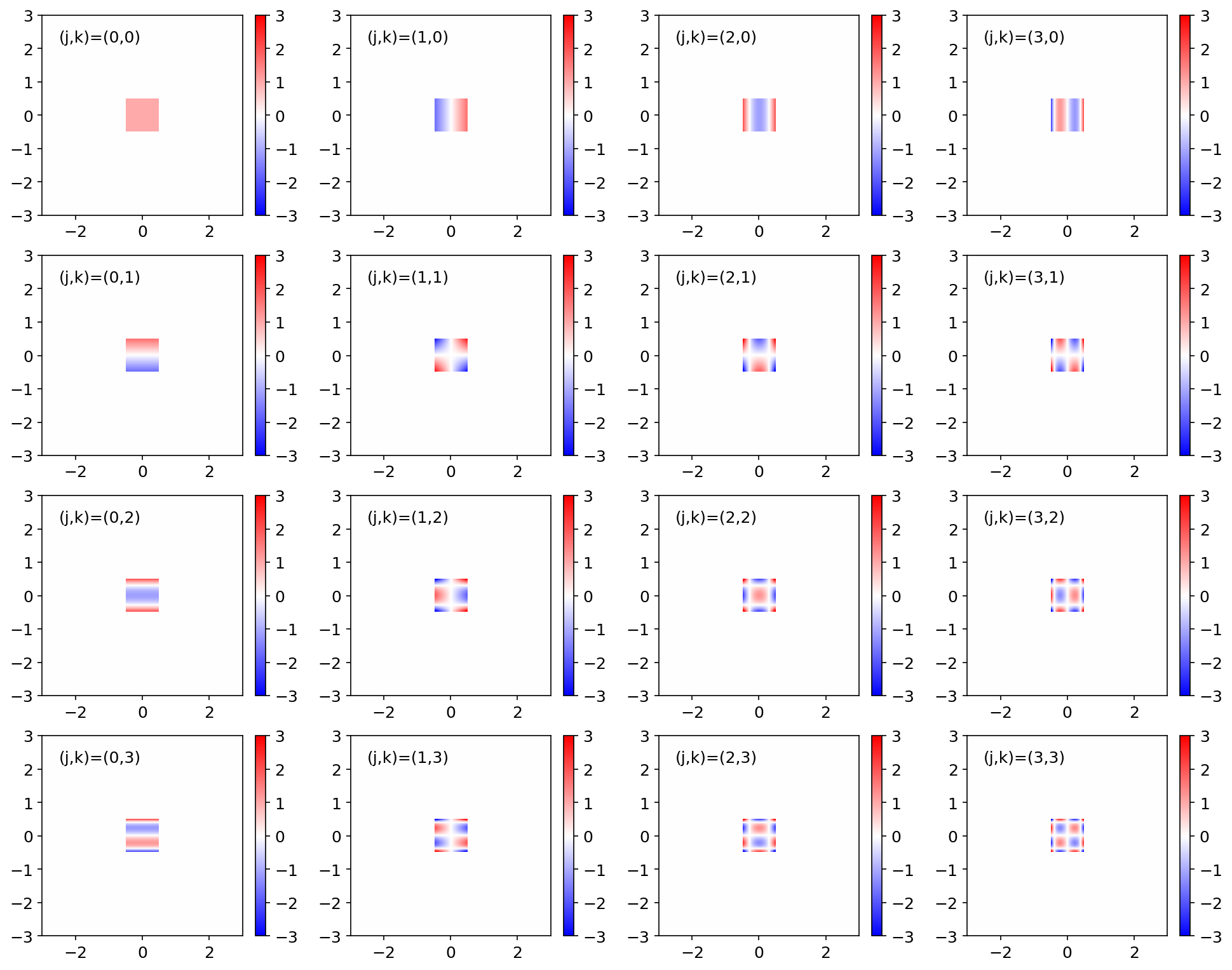}
    \caption{The orthonormal base functions $b_{jk}(\alpha,\beta)$ that we used to expand low-order wavefront errors ($ j,k, \leq 3$). }
    \label{fig:PB}
\end{figure}

\begin{figure}[h]
    \centering
    \includegraphics[width=1.0\linewidth]{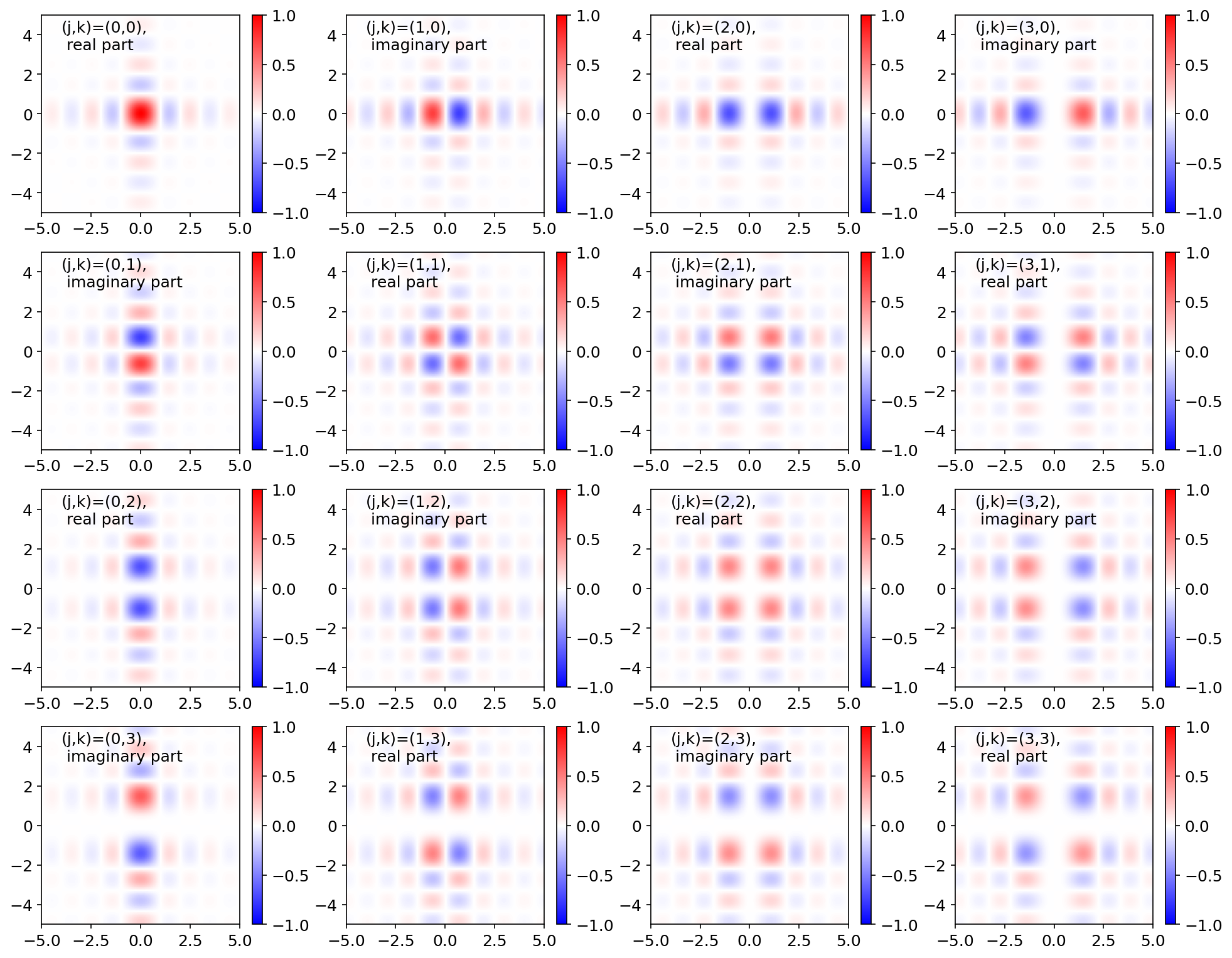}
    \caption{The Fourier conjugates of the base functions indicated in Figure \ref{fig:PB}. Since Fourier Transform preserves inner products of any two functions, these functions also have orthonormality.}
    \label{fig:FB}
\end{figure}

\begin{figure}[h]
    \centering
    \includegraphics[width=1.0\linewidth]{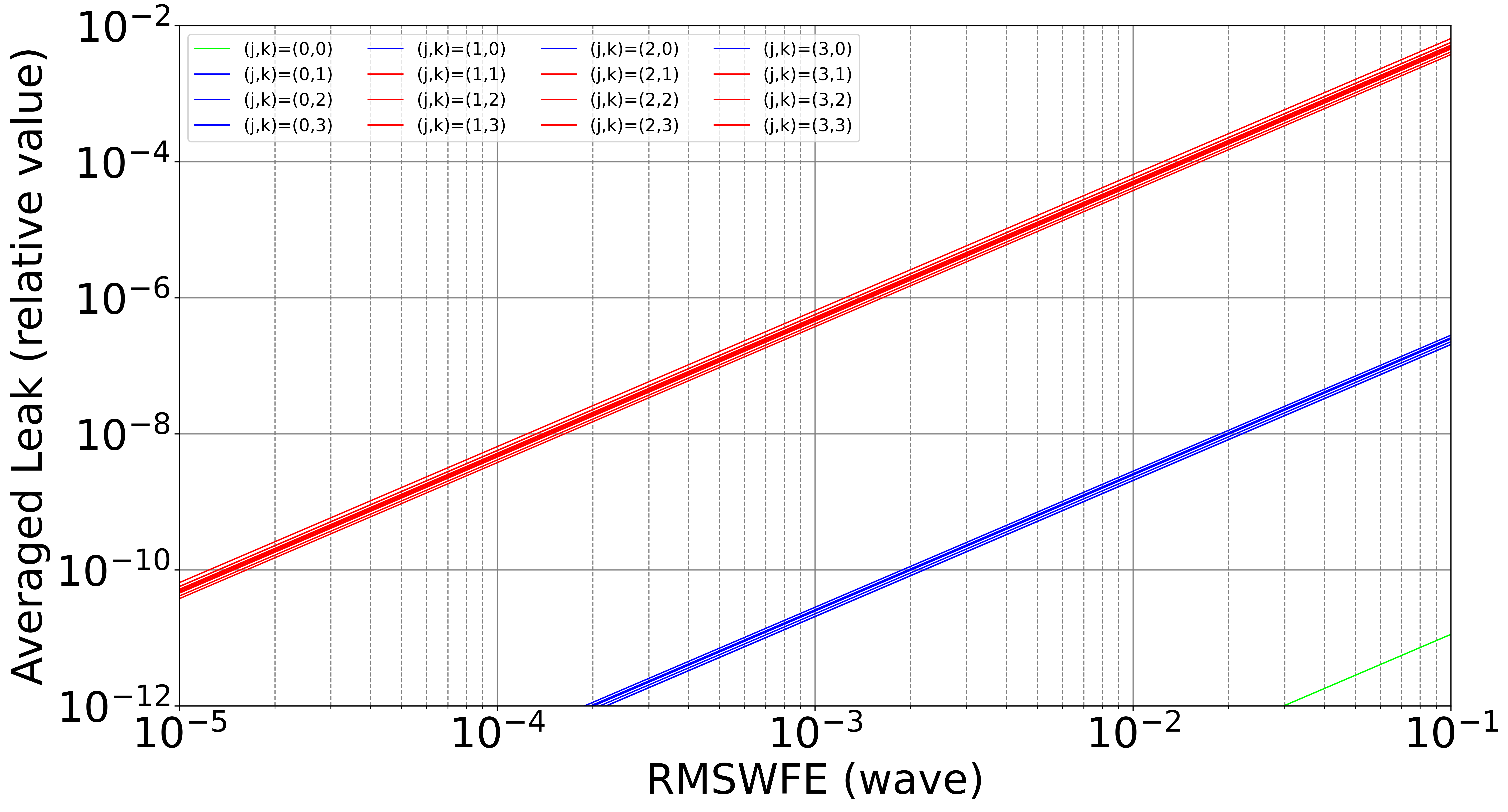}
    \caption{The relations between the magnitude of aberration coefficient $c_{jk}$ (waves in root mean square) of a particular order ($ j,k \leq 3$)  and the averaged leaks that the aberration brings. The simulation is executed for the wavelength of 690nm. The green solid curve indicates the case of $(j,k)=(0,0)$. The blue solid curves indicate the case in which one of ${j,k}$ is zero and another is not zero. The red solid curves indicate the case in which neither of ${j,k}$ is zero.  }
    \label{fig:AL}
\end{figure}

\begin{figure}[h]
    \centering
    \includegraphics[width=1.0\linewidth]{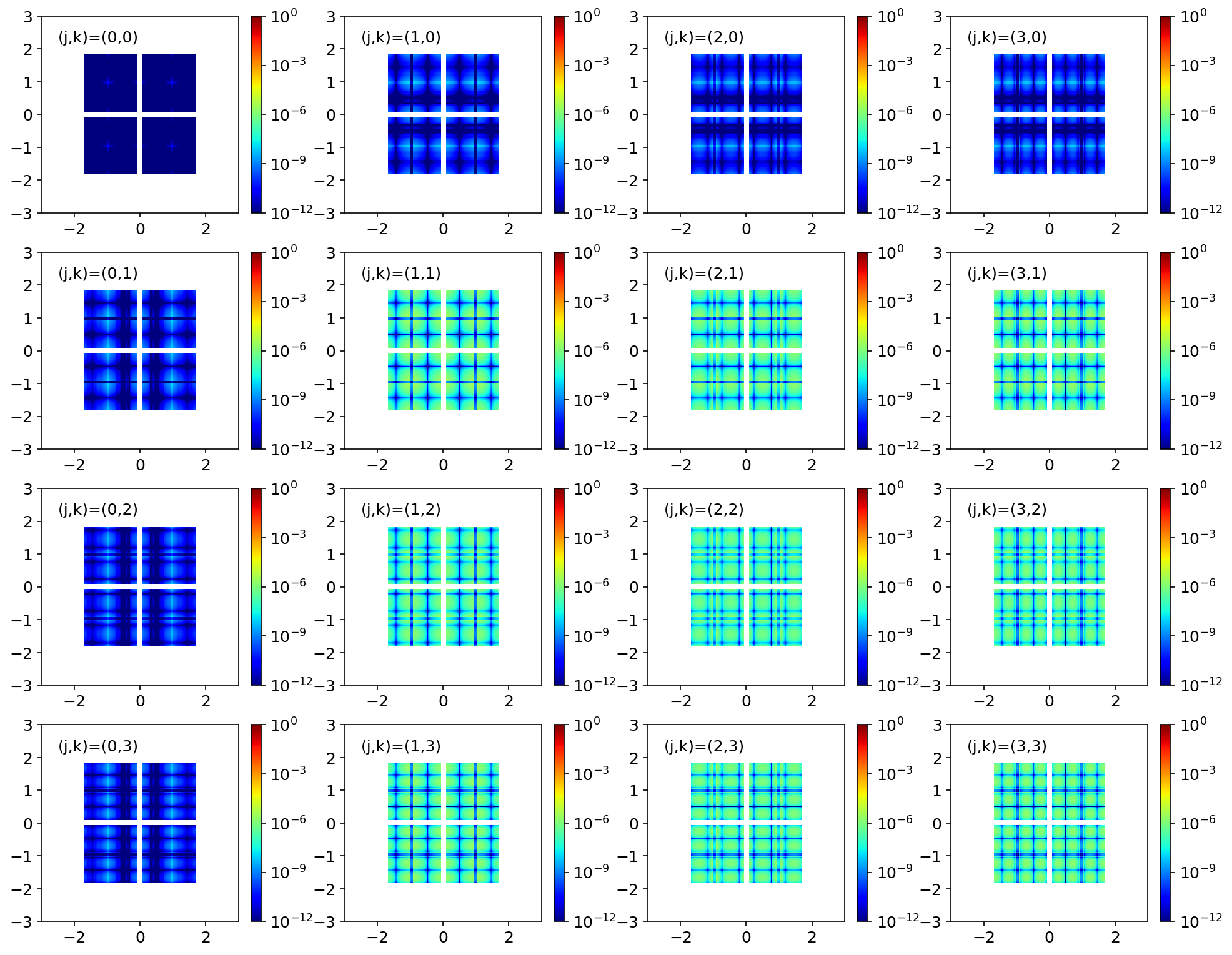}
    \caption{The leak intensity map with respect to each order of the aberration on the last Lyot stop. The magnitude of the aberration coefficient $c_{jk}$ is assumed to $10^{-3}$ waves. The assumed observational wavelength is 690nm.}
    \label{fig:PL}
\end{figure}

\begin{figure}[h]
    \centering
    \includegraphics[width=1.0\linewidth]{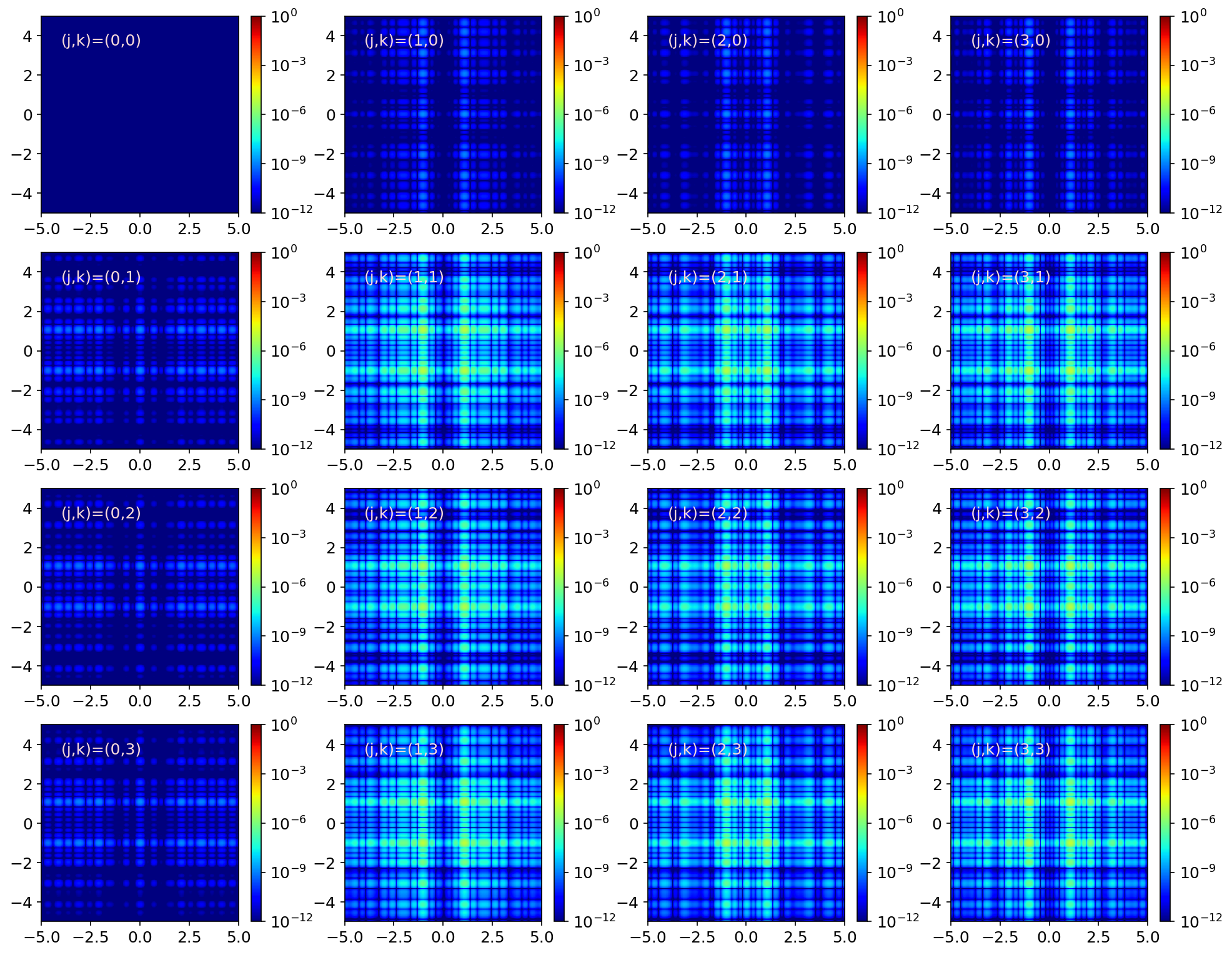}
    \caption{The leak intensity map with respect to each order of the aberration on the last focal plane. The magnitude of the aberration coefficient $c_{jk}$ is assumed to $10^{-3}$ waves. The assumed observational wavelength is 690nm.}
    \label{fig:FL}
\end{figure}

 { \section{Discussion}
 \subsection{Sensitivities to Low-order Aberrations}
 Sensitivities of leaks to low-order aberrations and misalighment of the coronagraphic components are also important performance criteria for a practical coronagraph.
Hence, we execute additional simulations of the sensitivities to low-order aberrations.
When the pupil function without aberration is $A(\alpha,\beta)$, the pupil function with aberration is  $A(\alpha,\beta)e^{i\Phi(\alpha,\beta)}$, where $\Phi(\alpha,\beta)$ denotes the wavefront-aberration function.
Here, we assume $\Phi(\alpha,\beta)=t\phi(\alpha,\beta)$, where $t$ denotes the magnitude of the aberration.
We can expand the exponential function of the phase factor $e^{i\Phi(\alpha,\beta)}$ as follows:
\begin{equation}
    e^{i\Phi(\alpha,\beta)}=1+it\phi(\alpha,\beta)-\frac{1}{2}t^2\phi(\alpha,\beta)^2+O\left(t^3\right).
    \label{eqet}
\end{equation}
The term of $t^n$ in Equation (\ref{eqet}) contributes the intensity leak of coronagraphs proportionally to $t^{2n}$.

In the simulations, we use scaled, normalized Legendre polynomials as base functions for expanding the wavefront-aberration function. 
The two-dimensional base functions $b_{jk}(\alpha,\beta)$ are defined as follows:
\begin{equation}
    b_{jk}(\alpha,\beta)=\sqrt{(2j+1)}P_{j}(2\alpha)\mathrm{rect}\left[\alpha\right]\sqrt{(2j+1)}P_{k}(2\beta)\mathrm{rect}\left[\beta\right],
\end{equation}
where $j$ and $k$ take non-negative integers and $P_{n}(z)$ is the Legendre polynomials of the $n$-th order.
Using the base functions $b_{jk}(\alpha ,\beta )$ and expansion coefficients $c_{jk}$, we can express $\Phi(\alpha,\beta)$ as following:
\begin{equation}
    \Phi(\alpha,\beta)=\sum_{j=0}^{\infty}\sum_{k=0}^{\infty}c_{jk}b_{jk}(\alpha ,\beta ).
\end{equation}
Figure \ref{fig:PB} shows the base functions $b_{jk}(\alpha,\beta)$ of the orders of $j,k \leq 3$;
the Fourier conjugates of these base functions are shown in Figure \ref{fig:FB}. 

The following orthonormality is a favorable property of defined base function $b_{jk}(\alpha,\beta)$:
\begin{equation}
    \int\!\!\!\!\int_{-\infty}^{\infty}\!\!\!\!d\alpha d\beta\  b_{jk}(\alpha,\beta)b_{lm}(\alpha,\beta)=\delta_{jl}\delta_{km},
\end{equation}
where $\delta_{nm}$ is the Kronecker delta.
From this property, when we can expand a wavefront error $\Psi(\alpha,\beta)$ as $\Psi(\alpha,\beta)=\sum_{j=0}^{\infty}\sum_{k=0}^{\infty}c_{jk}b_{jk}(\alpha,\beta)$, we can evaluate the root-mean-square wavefront error (RMSWFE) with the following relation:
\begin{equation}
    \mathrm{RMSWFE}=\sqrt{\sum_{j=1}^{\infty}\sum_{k=1}^{\infty}c_{jk}^2}.
\end{equation}
Hence, we can regard the coefficient $c_{jk}$ as a part of the RMSWFE contributed by the order $(j,k)$ that we are considering.

To investigate how much the leak is sensitive to the various low-order aberrations,  we simulated the averaged leaks (the same definition as the one in Section \ref{sec:sim}) that comes from only the coefficient $c_{jk}$ of a particular order for the case with various magnitudes of the aberrations. 
For this simulation, we assume the coronagraph system same as the one used in Section \ref{sec:sim} and observational wavelength of 690nm.
The results for the orders less than fourth are compiled in Figure \ref{fig:AL}.
From the results of the simulation, the orders $(j,k)$ are classified into the following three groups: (i) $(j,k)=(0,0)$, (ii) $(j,k)\neq(0,0)$ and $jk=0$, (iii) $jk \neq 0$. 
In Figure \ref{fig:AL}, the results of the cases included in groups (i), (ii) and (iii) are indicated as green, blue and red solid curves, respectively. 
For reference,  in Figure \ref{fig:PL} and \ref{fig:FL}, we show the leak-intensity maps on the last Lyot stop and the focal plane, respectively,  with respect to each order; in these figures, the aberration coefficient $c_{jk}$ is assumed to $10^{-3}$ waves (in root mean square). 

The base functions with orders of groups (i) and (ii) contribute to the leak in the considered system while they do not bring any second-order leaks in the coronagraph system with the mask function $M'_{\zeta}(x_{\lambda})$. However, the contribution is less problematic than that by group (iii); the leak due to group (iii) has approximately same magnitudes as the typical second-order aberrations (i.e. vortex coronagraphs with the topological charge of $l=2$ that has an IWA of 0.9 $\lambda/D$).
Reducing the wavefront error in group (iii) is an important challenge for us to practically use the proposed coronagraph system. 
}

\section{Conclusion}
In this paper, we describe an upgrade of a coronagraph system proposed by \citet{Itoh+2020} (the IM-2020 system) that enables it to be used over a moderate wavelength band. The upgrade is based on the fact that the wavelength-deviation leak produced at the Lyot stop is spatially constant.
An additional coronagraphic system reduces the leak at the pupil plane by the same mechanism as it reduces on-axis stellar light.
The development of this upgraded system required overcoming the problem of low off-axis throughput of a single IM-2020 system.

To solve this problem, we found that the off-axis throughput increases when the mask function includes a periodic factor that has the same zero-point interval as that of the side lobe of the original PSF.
Our performance simulation showed that a coronagraph system constructed with the new focal-plane mask approximately achieves a contrast below $10^{-10}$ over the spectral band of 650--750nm.
In addition, this coronagraph system achieves a core throughput greater than 10\% at the separation angles of 0.7--1.4$\lambda_0/D$; { To save observation time, the target's orbit must be known by other observational methods to a precision higher than the width of the field of view 0.7$\lambda_0/D$.}
For some types of aberration including tilt aberration, the proposed system has a sensitivity less than vortex coronagraphs ($l$=2) that has an IWA of $0.9\lambda/D$.
{ This type of coronagraphic system can be dedicated to realizing the extremely small IWAs. 
Using this system as a supplementary instrument could be a good solution for spectral characterizations of the atmospheres of Earth-like planets. }

\bibliography{bibbib}{}

\begin{thebibliography}{}
\expandafter\ifx\csname natexlab\endcsname\relax\def\natexlab#1{#1}\fi
\providecommand{\url}[1]{\href{#1}{#1}}
\providecommand{\dodoi}[1]{doi:~\href{http://doi.org/#1}{\nolinkurl{#1}}}
\providecommand{\doeprint}[1]{\href{http://ascl.net/#1}{\nolinkurl{http://ascl.net/#1}}}
\providecommand{\doarXiv}[1]{\href{https://arxiv.org/abs/#1}{\nolinkurl{https://arxiv.org/abs/#1}}}

\bibitem[{{Belikov} {et~al.}(2018){Belikov}, {Bryson}, {Sirbu}, {Guyon},
  {Bendek}, \& {Kern}}]{Belikov+2018}
{Belikov}, R., {Bryson}, S., {Sirbu}, D., {et~al.} 2018, in Society of
  Photo-Optical Instrumentation Engineers (SPIE) Conference Series, Vol. 10698,
  Space Telescopes and Instrumentation 2018: Optical, Infrared, and Millimeter
  Wave, 106981H, \dodoi{10.1117/12.2314202}

\bibitem[{{Benneke} {et~al.}(2019){Benneke}, {Wong}, {Piaulet}, {Knutson},
  {Lothringer}, {Morley}, {Crossfield}, {Gao}, {Greene}, {Dressing},
  {Dragomir}, {Howard}, {McCullough}, {Kempton}, {Fortney}, \&
  {Fraine}}]{Benneke+2019}
{Benneke}, B., {Wong}, I., {Piaulet}, C., {et~al.} 2019, \apjl, 887, L14,
  \dodoi{10.3847/2041-8213/ab59dc}

\bibitem[{{Cady} {et~al.}(2017){Cady}, {Balasubramanian}, {Gersh-Range},
  {Kasdin}, {Kern}, {Lam}, {Mejia Prada}, {Moody}, {Patterson}, {Poberezhskiy},
  {Riggs}, {Seo}, {Shi}, {Tang}, {Trauger}, {Zhou}, \& {Zimmerman}}]{Cady+2017}
{Cady}, E., {Balasubramanian}, K., {Gersh-Range}, J., {et~al.} 2017, in Society
  of Photo-Optical Instrumentation Engineers (SPIE) Conference Series, Vol.
  10400, Society of Photo-Optical Instrumentation Engineers (SPIE) Conference
  Series, 104000E, \dodoi{10.1117/12.2272834}

\bibitem[{{Des Marais} {et~al.}(2002){Des Marais}, {Harwit}, {Jucks},
  {Kasting}, {Lin}, {Lunine}, {Schneider}, {Seager}, {Traub}, \&
  {Woolf}}]{DesMarais+2002}
{Des Marais}, D.~J., {Harwit}, M.~O., {Jucks}, K.~W., {et~al.} 2002,
  Astrobiology, 2, 153, \dodoi{10.1089/15311070260192246}

\bibitem[{{Fujii} {et~al.}(2018){Fujii}, {Angerhausen}, {Deitrick},
  {Domagal-Goldman}, {Grenfell}, {Hori}, {Kane}, {Pall{\'e}}, {Rauer},
  {Siegler}, {Stapelfeldt}, \& {Stevenson}}]{Fujii+2018}
{Fujii}, Y., {Angerhausen}, D., {Deitrick}, R., {et~al.} 2018, Astrobiology,
  18, 739, \dodoi{10.1089/ast.2017.1733}

\bibitem[{{Guyon} {et~al.}(2014){Guyon}, {Hinz}, {Cady}, {Belikov}, \&
  {Martinache}}]{Guyon+2014}
{Guyon}, O., {Hinz}, P.~M., {Cady}, E., {Belikov}, R., \& {Martinache}, F.
  2014, \apj, 780, 171, \dodoi{10.1088/0004-637X/780/2/171}

\bibitem[{{Guyon} {et~al.}(2010){Guyon}, {Martinache}, {Belikov}, \&
  {Soummer}}]{Guyon+2010}
{Guyon}, O., {Martinache}, F., {Belikov}, R., \& {Soummer}, R. 2010, \apjs,
  190, 220, \dodoi{10.1088/0067-0049/190/2/220}

\bibitem[{{Itoh} \& {Matsuo}(2020)}]{Itoh+2020}
{Itoh}, S., \& {Matsuo}, T. 2020, \aj, 159, 213,
  \dodoi{10.3847/1538-3881/ab811c}

\bibitem[{{Kaltenegger}(2017)}]{Kaltenegger+2017}
{Kaltenegger}, L. 2017, \araa, 55, 433,
  \dodoi{10.1146/annurev-astro-082214-122238}

\bibitem[{{Kasting} {et~al.}(2009){Kasting}, {Traub}, {Roberge}, {Leger},
  {Schwartz}, {Wootten}, {Vosteen}, {Lo}, {Brack}, {Tanner}, {Coustenis},
  {Lane}, {Oppenheimer}, {Mennesson}, {Lopez}, {Grillmair}, {Beichman},
  {Cockell}, {Hanot}, {McCarthy}, {Stark}, {Marois}, {Aime}, {Angerhausen},
  {Montes}, {Wilner}, {Defrere}, {Mourard}, {Lin}, {Kite}, {Chassefiere},
  {Malbet}, {Tian}, {Westall}, {Illingworth}, {Vasisht}, {Serabyn}, {Marcy},
  {Bryden}, {White}, {Laughlin}, {Torres}, {Hammel}, {Ferguson}, {Shibai},
  {Rottgering}, {Surdej}, {Wiseman}, {Ge}, {Bally}, {Krist}, {Monnier},
  {Trauger}, {Horner}, {Catanzarite}, {Harrington}, {Nishikawa}, {Stapelfeldt},
  {von Braun}, {Biazzo}, {Carpenter}, {Balasubramanian}, {Kaltenegger},
  {Postman}, {Spaans}, {Turnbull}, {Levine}, {Burchell}, {Ealey}, {Kuchner},
  {Marley}, {Dominik}, {Mountain}, {Kenworthy}, {Muterspaugh}, {Shao}, {Zhao},
  {Tamura}, {Kasdin}, {Haghighipour}, {Kiang}, {Elias}, {Woolf}, {Mason},
  {Absil}, {Guyon}, {Lay}, {Borde}, {Fouque}, {Kalas}, {Lowrance}, {Plavchan},
  {Hinz}, {Kervella}, {Chen}, {Akeson}, {Soummer}, {Waters}, {Barry},
  {Kendrick}, {Brown}, {Vanderbei}, {Woodruff}, {Danner}, {Allen}, {Polidan},
  {Seager}, {MacPhee}, {Hosseini}, {Metchev}, {Kafka}, {Ridgway}, {Rinehart},
  {Unwin}, {Shaklan}, {ten Brummelaar}, {Mazeh}, {Meadows}, {Weiss}, {Danchi},
  {Ip}, \& {Rabbia}}]{Kasting+2009}
{Kasting}, J., {Traub}, W., {Roberge}, A., {et~al.} 2009, in astro2010: The
  Astronomy and Astrophysics Decadal Survey, Vol. 2010, 151.
\newblock \doarXiv{0911.2936}

\bibitem[{{Matsuo} {et~al.}(2021){Matsuo}, {Itoh}, \& {Ikeda}}]{Matsuo+2021}
{Matsuo}, T., {Itoh}, S., \& {Ikeda}, Y. 2021, \aj, 161, 83,
  \dodoi{10.3847/1538-3881/abd248}

\bibitem[{{Mawet} {et~al.}(2011){Mawet}, {Serabyn}, {Wallace}, \&
  {Pueyo}}]{Mawet+2011a}
{Mawet}, D., {Serabyn}, E., {Wallace}, J.~K., \& {Pueyo}, L. 2011, Optics
  Letters, 36, 1506, \dodoi{10.1364/OL.36.001506}

\bibitem[{{N'Diaye} {et~al.}(2016){N'Diaye}, {Soummer}, {Pueyo}, {Carlotti},
  {Stark}, \& {Perrin}}]{N'Diaye+2016}
{N'Diaye}, M., {Soummer}, R., {Pueyo}, L., {et~al.} 2016, \apj, 818, 163,
  \dodoi{10.3847/0004-637X/818/2/163}

\bibitem[{{Pueyo} \& {Norman}(2013)}]{Pueyo+2013}
{Pueyo}, L., \& {Norman}, C. 2013, \apj, 769, 102,
  \dodoi{10.1088/0004-637X/769/2/102}

\bibitem[{{Ruane} {et~al.}(2018){Ruane}, {Mawet}, {Mennesson}, {Jewell}, \&
  {Shaklan}}]{2018JATIS...4a5004R}
{Ruane}, G., {Mawet}, D., {Mennesson}, B., {Jewell}, J., \& {Shaklan}, S. 2018,
  Journal of Astronomical Telescopes, Instruments, and Systems, 4, 015004,
  \dodoi{10.1117/1.JATIS.4.1.015004}

\bibitem[{{Seager} {et~al.}(2016){Seager}, {Bains}, \&
  {Petkowski}}]{Seager+2016}
{Seager}, S., {Bains}, W., \& {Petkowski}, J.~J. 2016, Astrobiology, 16, 465,
  \dodoi{10.1089/ast.2015.1404}

\bibitem[{{Stark} {et~al.}(2019){Stark}, {Belikov}, {Bolcar}, {Cady}, {Crill},
  {Ertel}, {Groff}, {Hildebrandt}, {Krist}, {Lisman}, {Mazoyer}, {Mennesson},
  {Nemati}, {Pueyo}, {Rauscher}, {Riggs}, {Ruane}, {Shaklan}, {Sirbu},
  {Soummer}, {Laurent}, \& {Zimmerman}}]{Stark+2019}
{Stark}, C.~C., {Belikov}, R., {Bolcar}, M.~R., {et~al.} 2019, Journal of
  Astronomical Telescopes, Instruments, and Systems, 5, 024009,
  \dodoi{10.1117/1.JATIS.5.2.024009}

\bibitem[{{Trauger} {et~al.}(2012){Trauger}, {Moody}, {Gordon}, {Krist}, \&
  {Mawet}}]{Trauger+2012}
{Trauger}, J., {Moody}, D., {Gordon}, B., {Krist}, J., \& {Mawet}, D. 2012, in
  Society of Photo-Optical Instrumentation Engineers (SPIE) Conference Series,
  Vol. 8442, Space Telescopes and Instrumentation 2012: Optical, Infrared, and
  Millimeter Wave, 84424Q, \dodoi{10.1117/12.926663}

\bibitem[{{Tsiaras} {et~al.}(2019){Tsiaras}, {Waldmann}, {Tinetti}, {Tennyson},
  \& {Yurchenko}}]{Tsiaras+2019}
{Tsiaras}, A., {Waldmann}, I.~P., {Tinetti}, G., {Tennyson}, J., \&
  {Yurchenko}, S.~N. 2019, Nature Astronomy, 3, 1086,
  \dodoi{10.1038/s41550-019-0878-9}

\end{thebibliography}
\bibliographystyle{aasjournal}

\appendix
\section{Effects of the Cosine Modulation}
Here, we assume $\lambda=\lambda_0$ for simplicity.
The mask function $M'_{\zeta}(x_{\lambda_0})$ modulates the ASF of an on-axis point source $\tilde{P}(x_{\lambda_0})$ as follows:
\begin{eqnarray}
    M'_{\zeta}(x_{\lambda_0})\tilde{P}(x_{\lambda_0})&=&C_{\zeta}'\left(1-2\zeta^{-1}\sinc(4x_{\lambda_0})\right)\cos(\pi x_{\lambda_0})\sinc(x_{\lambda_0})\nonumber\\
    &=&C_{\zeta}'\left(1-2\zeta^{-1}\sinc(4x_{\lambda_0})\right)\sinc(2x_{\lambda_0}),
    \label{e4}
\end{eqnarray}
where we use the mathematical formula:
\begin{equation}
\sin(2A)=2\sin(A)\cos(A).
\end{equation}
From Equation (\ref{e4}), we observe  that $M'_{\zeta}(x_{\lambda_0})\tilde{P}(x_{\lambda_0})$ is proportional to $M(2x_{\lambda_0})\tilde{P}(2x_{\lambda_0})$. Consequently, $M'_{\zeta}(x_{\lambda_0})$ behaves similarly to $M_{\zeta}(x_{\lambda_0})$ for an on-axis point source (i.e., it produces a perfect null) although the size of the Lyot stop doubles compared to the case with $M_{\zeta}(x_{\lambda_0})$. 

The mask function $M'_{\zeta}(x_{\lambda_0})$ modulates the ASF of point sources that have separation angles of the integer $N$ times $\lambda_0/D$  as follows:
\begin{eqnarray}
    M'_{\zeta}(x_{\lambda_0})\tilde{P}'(x_{\lambda_0}-N)    &=&(C_{\zeta}'/C_{\zeta})(-1)^{N}(M(2x_{\lambda_0})\tilde{P}(2(x_{\lambda_0}-N)).
    \label{e7}
\end{eqnarray}
Hence, $M'_{\zeta}(x_{\lambda_0})$ works as $M_{\zeta}(x_{\lambda_0})$ for point sources with separation angles of  $N\lambda_0/D$. The width of the ASF at the next focal plane is halved.

{ We should clearly distinguish the cosine factor's doubling the size of the re-imaged pupil from the usual magnification of re-imaging the pupil. In first-order geometrical optics, when the pupil is magnified by a factor of $\gamma$, the focal image is magnified by a factor of $\gamma^{-1}$; we use this relation to normalize the coordinates on the pupil and focal plains in wave-optical calculation so that the scales that are optically conjugated to each other have same values of the coordinates. 
However, the cosine factor that doubles the size of the re-imaged pupil does not change the focal-image magnification while usual magnification of pupils inevitably change the focal-image magnification. 
Hence, in normalization of the coordinates, we need to treat the enlargement of the size of the re-imaged pupil by the cosine factor in a different way from the usual magnification of re-imaging the pupil.
To be more specific, when $D$ and $\lambda/D$ normalize the pupil- and focal- plane coordinates, this $D$ is not be affected by the cosine modulation. 
}

\begin{figure}[h]
    \centering
    \includegraphics[width=1.0\linewidth]{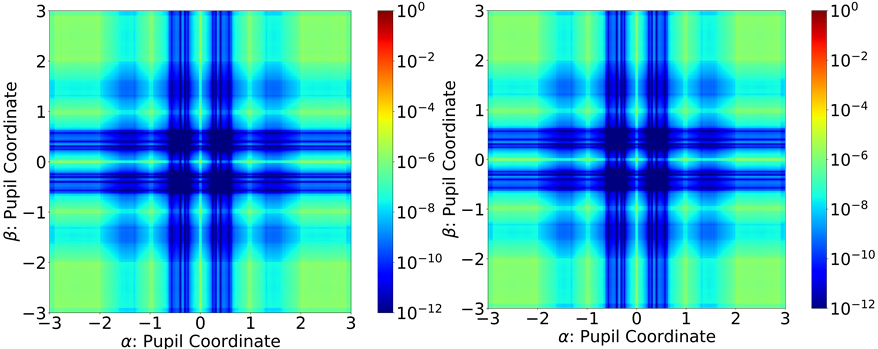}
    \caption{Comparison of the results of numerical calculations of pupil leaks  with different discretization: the result with the same mesh to discretize the focal plane as the one used in Section \ref{sec:sim} (left) and the result with the mesh 16-times finer than the one used in Section \ref{sec:sim} (right). The assumed observational wavelength is 700nm.}
    \label{fig:com}
\end{figure}

{
\section{Numerical Fourier Transform and its Accuracy}
When the pupil- and focal plane coordinates are normalized by $D$ and $\lambda/D$, the amplitude on these planes are related to each other in the following way:
\begin{equation}
    \tilde{f}(x)=\int_{-\infty}^{\infty}d\alpha f(\alpha) e^{-2\pi i x \alpha},
\label{ApB1}
\end{equation}
where we extracted one-dimensional Fourier transform from two-dimensional one for simplicity.  
For numerically approximating the Fourier transform, we used the DFT method that is defined as follows:
\begin{equation}
    \tilde{a}_{l}=\mathrm{DFT}\left[a_{k}\right]=\sum_{k=0}^{N-1} a_{k} e^{-\frac{2\pi i l k}{N}},
\label{ApB2}
\end{equation}
where $N$ is the number of array elements used for the DFT.
To rewrite Equation (\ref{ApB1}) into the form of Equation (\ref{ApB2}), we sampled the coordinates $x$ and $\alpha$ using the sampling intervals $\Delta x$ and $\Delta \alpha$ as follows:
\begin{equation}
    x_{l}=\Delta x \left\lbrace l-\frac{N-1}{2}\right\rbrace
   \label{ApB3} 
\end{equation}
and
\begin{equation}
    \alpha_{k}=\Delta \alpha \left\lbrace k-\frac{N-1}{2}\right\rbrace,
    \label{ApB4}
\end{equation}
where k and l are integers and we assume that $N$ is an even number; and the pupil- and focal sampling intervals must satisfy the following equation:
\begin{equation}
    \Delta x \Delta \alpha = 1/N.
\end{equation}
After substituting Equation (\ref{ApB3}) and (\ref{ApB4}), we obtain an approximation of $\tilde{f}(x_{l})$:
\begin{equation}
   \tilde{f}(x_{l})\approx  \tilde{f_{\mathrm{ap}}}(x_{l}) = i^{\frac{(N-1)^2}{N}} \Delta x \Upsilon_{l}^{N} \mathrm{DFT}\left[\Upsilon_{k}^{N} f(\alpha_{k})\right],
\label{ApB5}
\end{equation}
where we defined $\Upsilon_{l}^{N}$ in the following way:
\begin{equation}
    \Upsilon_{l}^{N}= (-1)^{\left\lbrace l\left(1-\frac{1}{N}\right)\right\rbrace}.
\end{equation}
We used Equation (\ref{ApB5}) to execute all the one-dimensional numerical Fourier transforms in this study; all the two-dimensional Fourier transforms that appear in this study can be resolved into one-dimensional numerical Fourier transforms.   

To avoid aliasing noises, we adopt the following discretization parameters:
\begin{eqnarray}
    N&=&1048576 \\
    \Delta x &=&\frac{1}{40.96} \\
    \Delta \alpha &=&\frac{1}{25600}.
\end{eqnarray}
When the pupil sampling interval $\Delta \alpha$ is $\frac{1}{25600}$, the sinusoidal-modulation  components on the pupil plane with the spatial-frequency differences of integer times 25600 are indistinguishable from each other. 
Thus, the approximation of its Fourier transform $\tilde{f_{\mathrm{ap}}}(x_{l})$ has periodicity as follows:
\begin{equation}
    \tilde{f_{\mathrm{ap}}}(x_{l})= \sum_{n=-\infty}^{\infty} \tilde{f}(x_{l}-25600 n).
\label{ApB6}
\end{equation}
In Equation (\ref{ApB6}), the terms of $n\neq 0$ contributes numerical errors as aliasing noises.  
Since only the terms with small $|n|$ have great impact on the error in the region near the origin of the focal coordinates, Equation (\ref{ApB6}) can be practically simplified as follows:
\begin{equation}
    \tilde{f_{\mathrm{ap}}}(x_{l})=\tilde{f}(x_{l})+ \tilde{f}(x_{l}-25600)+\tilde{f}(x_{l}+25600),
\label{ApB7}
\end{equation}
where we omitted the terms of $|n|\leq 2 $.
For example, when we assume $\tilde{f}(x)=\mathrm{sinc}(x)$ as a typical function on the focal planes, the absolute value of $\tilde{f}(x)$ is less than $\frac{1}{\pi x}$. Thus, the total contribution by second and third terms of Equation (\ref{ApB7}) to the aliasing noise at $x_l\approx 0$ is approximately less than $\frac{2}{25600\pi}\approx 2.5\times 10^{-5}$ in amplitude (i.e. $6.2 \times 10^{-10}$ in intensity).
The estimation of the amount of the aliasing noise is not largely altered even when we execute two-dimensional Fourier transform by multiplying two one-dimensional transforms.
This is due to existence of the cross term between an error term of a dimension and a true-value term of another dimension. 
Hence, for the calculation from pupil planes to focal planes, we estimated the upper limit of the relative numerical error with respect to the input pupil intensity as $6.2 \times 10^{-10}$.

To move the topic to the calculation from focal plane to pupil plane, all we have to do is to exchange the definitions of the pupil- and focal coordinate symbols. 
After this exchange, the approximation of the Fourier transform $\tilde{f_{\mathrm{ap}}}(x_{l})$ has a period of 40.96 due to the aliasing noise. 
However, here, the aliasing is not problematic because functions on the pupil plane usually have non-zero values only at a region near the origin (i.e. $\mathrm{rect}(\alpha)=0$ for $|\alpha|>0.5$).  
In fact, the aliasing noise has no impact on the focus-to-pupil calculation of the coronagraph system based on the mask function $M'_{\zeta}(x_{\lambda})$ [Equation (\ref{e1})].

However, the mask function $M''_{\zeta}(G,H;x_{\lambda})$ [Equation (\ref{e2})] may include some high-(spatial)-frequency components that might contribute to the numerical error. 
To test this contribution, we compared the result of calculations with the two different sampling intervals using the final pupil leaks of the system as a test case.
Figure \ref{fig:com} includes two panels: the case with the same sampling interval as the one used in Section \ref{sec:sim} (left) and the case with the sampling interval 16-times finer than the one used in Section \ref{sec:sim} (right). 
From these results, we expect that the focus-to-pupil calculation in this study has no critical numerical errors.

}

\end{document}